\documentclass[aps,twocolumn,showpacs,floats]{revtex4}
\usepackage{graphicx}

\begin{document}

\title{Nucleon-Meson Couplings in One Boson Exchange Potential
using the Non-Critical String Theory}

\author{M.\ R.\ Pahlavani
\footnote{m.pahlavani@umz.ac.ir}
       }

\author{J.\ Sadeghi
\footnote{pouriya@ipm.ir}
       }
\author{R.\ Morad
\footnote{r.morad@umz.ac.ir}
       }

\affiliation{
Department of Nuclear Physics, Faculty of Basic science, University of Mazandaran, P.O.Box 47416-1467, Babolsar, Iran\\
             }
\begin{abstract}
A non-critical holographic QCD model constructed in the six
dimensional Anti-de-Sitter ($AdS_6$) supergravity background is
employed to study a the baryon. It is shown that the size of the
baryon is of order one with respect to the $\lambda$, however, it is
smaller than the scale of the dual QCD. An effective four
dimensional action for the nucleon is obtained in terms of the meson
exchange potentials. All meson-nucleon couplings in the non-critical
$AdS_6$ background are calculated. Results obtained using our model
are compared with predictions of four modern phenomenological
interaction models. Also, our numerical results are compared with
the results of the Sakai-Sugimoto (SS) model which indicate that the
non-critical holographic QCD model can be a good toy to calculate
the meson-nucleon couplings.

\end{abstract}


\maketitle
\nonstopmode

\section{Introduction}

The construction of a nucleon-nucleon (NN) potential has a long
history in nuclear physics due to its role in understanding the
nuclear force. Many potential models have been constructed from the
1950s which have been composed to fit the available NN scattering
data. The newer potentials have only slightly improved with respect
to the previous ones in describing the recent much more accurate
data. As it is shown in Ref[1], all of these potential models do not
have good quality with respect to the pp scattering data below 350
MeV and just a few of them are of satisfactory quality. These models
are the Reid soft-core potential Reid68 [2], the Nijmegen soft-core
potential Nijm78 [3], the new Bonn pp potential Bonn89 [4] and also
the parameterized Paris potential Paris80 [5]. These familiar
one-boson-exchange potentials (OBEP) contain a relatively small
number of free parameters (about 10 to 15 parameters), but do not
have a reasonable description of the empirical scattering data.
Also, most of these potentials which have been fitted to the np
scattering data, unfortunately do not automatically fit to the pp
scattering data even by considering the correction term for the
Coulomb interaction [1]. Of course, new versions of these potentials
have been constructed such as Nijm I, Nijm II, Reid93 [6], CD-Bonn
[7], and AV18 [8] which explain the empirical scattering data
successfully. But they contain a large number of purely
phenomenological parameters. For example, an updated (Nijm92pp [9])
version of the Nijm78 potential contains 39 free parameters.

On the other hand, there are many attempts to impose the symmetries
of QCD using an effective Lagrangian of pions and nucleons [10,11].
These models only capture the qualitative features of the nuclear
interactions and could not compete with the much more successful
potential models mentioned above. Despite many efforts, no potential
model has not yet been constructed which gives a high-quality
description of the empirical data, obeys the symmetries of QCD, and
contains only a few number of free phenomenological parameters.

One of the applications of anti–de Sitter space/conformal field
theory (AdS/CFT) duality [12-14] is a holographic QCD introduced
recently to solve the strong-coupling QCD problems such as the
chiral dynamics of hadrons in particular baryons [15-38]. The
Sakai-Sugimoto (SS) [39, 40] and Klebanov-Strassler (KS) models [41]
are the most interesting holographic models.

The predictions of the SS model are in a good agreement with the
lattice simulations such as a glueball spectrum of pure QCD [42,
43]. Also this model describes baryons and their interactions with
mesons [22-24, 39, 40]. It is shown that the baryons can be taken as
 point-like objects at distances larger than their sizes, so
their interactions can be described by the exchange of light
particles such as mesons. Therefore, one can find the baryon-baryon
potential from the Feynman diagrams using the interaction vertices
including baryon currents and light mesons [23]. But there are some
inconsistencies. For example, the size of the baryon is proportional
to $\lambda^{-1/2}$. Consequently in the large 't Hooft coupling
(large $\lambda$), the size of the baryon becomes zero and the
stringy corrections have to be taken into account. Another problem
is that the scale of the system associated with the baryonic
structure is roughly half the one needed to fit to the mesonic data
[44].

Also there is another problem for such holographic models arising
from critical string theory. In these models, the color brane
backgrounds are ten-dimensional so the dual gauge theories are
supersymmetric. In order to break the supersymmetry, some parts of
such backgrounds need to be compacted on some manifolds. This causes
the production of some Kaluza-Klein (KK) modes, with the mass scale
of the same order as the masses of the hadronic modes. These
unwanted modes are coupled to the hadronic modes and there is no
mechanism to disentangle them from the hadronic modes yet. In order
to overcome this problem, it is possible to consider the color brane
configuration in non-critical string theory. The result is a
gravitational background located at the low dimensions [45-48]. In
this background the string coupling constant is proportional to
$\frac{1}{N_c}$, so the large $N_c$ limit corresponds to the small
string coupling constant. However, contrary to the critical
holographic models, in the large $N_c$ limit, the 't Hooft coupling
is of order one instead of infinity and the scalar curvature of the
gravitational background is also of order one. So, it seems the
non-critical gauge-gravity correspondence is not very reliable. But
studies show that the results of these models for some low energy
QCD properties such as the meson mass spectrum, Wilson loop, and the
mass spectrum of glueballs [49-51] are comparable with lattice
computations. Therefore non-critical holographic models still seem
useful to study QCD.

One of the non-critical holographic models is composed of a $D4$ and
anti $D4$ brane in six-dimensional non-critical string theory
[47,49]. The low energy effective theory on the intersecting brane
configuration is a four-dimensional QCD-like effective theory with
the global chiral symmetry $U(N_f)_L \times U(N_f)_R$. In this brane
configuration, the six-dimensional gravity background is the near
horizon geometry of the color $D4$ branes. This model is based on
the compactified $AdS_6$ space-time with constant dilaton. So the
model does not suffer from the large string coupling as the SS
model. The meson spectrum [49] and the structure of thermal phase
[52] are studied in this model. Some properties, like the dependence
of the meson masses on the stringy mass of the quarks and the
excitation number are different from the critical holographic models
such as the SS model.

In this paper, we are going to obtain the NN potential using the
non-critical $AdS_6$ background. We study the gauge field and its
mode expansion in this non-critical holography model and obtain the
pion action. The model has a mass scale $M_{KK}$ like the SS model
in which we set its value by computing the pion decay constant.
Then, we study the baryon and obtain its size. We show that the size
of the baryon is of order one with respect to the 't Hooft coupling,
so the problem of the zero size of the baryon in the critical
holography model is solved. But the size of the baryon is still
smaller than the mass scale of holographic QCD, so we treat it as a
point-like object and introduce an isospin $1/2$ Dirac field for the
baryon. We write a 5D effective action for the baryon field and
reduce it to the 4D using the mode expansion of gauge field and
baryon field and obtain the NN potential in terms of the meson
exchange interactions. We calculate the meson-nucleon couplings
using the suitable overlapping wave function integrals and compare
them with the results of SS model. Also, our results are compared
with predictions of some phenomenological models and also the SS
models for the couplings. Our study shows that the non-critical
results are in good agreement with the other available models.

This paper is organized as follows: In Sec. 2 we briefly review the
non-critical model and mode expansion of the gauge field. We analyze
the baryon and extract its mass and size in Sec. 3. In Sec. 4, an
effective action for the baryon is considered and a non-critical
prescription of nucleon-nucleon potential in terms of the meson
exchange interactions is obtained. In Sec. 5, the nucleon-meson
couplings are calculated and compared with predictions of four
modern phenomenological models (Nijmegen (93), Paris, CD-Bonn and AV
18 models). Section 6 is devoted to a brief summary and conclusions.

\section{  Holographic QCD from the non-critical string theory }

In the presented non-critical model, the gravity background is
generated by near-extremal $D4$ branes wrapped over a circle with
the anti-periodic boundary conditions. Two stacks of flavor branes,
namely $D4$ branes and anti-$D4$ branes, are added to this geometry
and are called flavor probe branes. The color branes extend along
the directions $t,x_1,x_2,x_3,$ and $\tau$ while the probe flavor
branes fill the whole Minkowski space and stretch along the radius
$U$ which is extended to infinity. The strings attaching a color
D4-brane to a flavor brane transform as quarks, while strings
hanging between a color $D4$ and a flavor $\overline{D4}$ transform
as anti-quarks. The chiral symmetry breaking is achieved by a
reconnection of the brane, anti-brane pairs. Under the quenched
approximation $(N_c \gg N_f)$, the reactions of flavor branes and
the color branes can be neglected. Just like the SS model, the
$\tau$ coordinate is wrapped on a circle and the anti-periodic
condition is considered for the fermions on the thermal circle. The
final low energy effective theory on the background is a
four-dimensional QCD-like effective theory with the global chiral
symmetry $U(N_f)_L\times U(N_f)_R$.

In this model, the near horizon gravity background at low energy is [49]
\begin{eqnarray}
ds^2&=&\left( \frac{U}{R} \right)^2 (-dt^2+dx_i dx_i+f(U) d\tau^2)
+ \nonumber\\
&&\left( \frac{R}{U} \right)^2 \frac{dU^2}{f(U)},
\end{eqnarray}
where R is the radius of the AdS space. Also $f(U)$ and RR
six-form field strength, $F_{(6)}$ are defined by the following
relations
\begin{eqnarray}
F_{(6)}&=&Q_c \left( \frac{U}{R} \right)^4 dt \wedge
dx_1\wedge dx_2 \wedge dx_3  \wedge du \wedge d\tau \,, \nonumber\\
f(U)&=&1-\left( \frac{U_{KK}}{U} \right)^5.
\end{eqnarray}
In order to obtain solutions of near extremal flavored $AdS_6$,
the values of dilaton and $R_{AdS}$ are considered as
\begin{eqnarray}
e^\phi &=& \frac{2}{3}\frac{Q_f}{Q_c^2}(\sqrt{1+\frac{6Q_c^2}{Q_f^2}}-1)\, ,\nonumber\\
R_{AdS}^2&=&\frac{90}{12+\frac{Q_f^2}{Q_c^2}-\frac{Q_f^2}{Q_c^2}\sqrt{1+\frac{6Q_c^2}{Q_f^2}}}.
\end{eqnarray}
This relation indicates that the $R_{AdS}$ and dilaton depend on the
ratio of the number of colors $(\sim Q_c)$ and flavors$(\sim Q_f)$.
Under the quenched approximation, the values of the dilaton and AdS
radius can be rewritten as,
\begin{eqnarray}
R_{AdS}^2=\frac{15}{2}\,\,\,\,,\,\,\,\,e^\phi &=
&\frac{2\sqrt2}{\sqrt3 Q_c},
\end{eqnarray}
where  $Q_c$ is proportional to the number of color branes, $N_c$.

To avoid singularity, the coordinate $\tau$ satisfies the following
periodic condition,
\begin{eqnarray}
\tau \sim \tau + \delta \tau\,\,,\qquad\qquad
\delta\tau=\frac{4\pi R^2}{5 U_{KK}}\,\,.
\end{eqnarray}
Also, the Kaluza-Klein mass scale of this compact dimension is
\begin{eqnarray}
M_{KK}=\frac{2\pi}{\delta\tau}=
\frac{5}{2} \frac{ \ U_{KK}}{R^2},
\end{eqnarray}
and dual gauge field theory for this background is non
supersymmetric. Also, the Yang-Mills coupling constants can be
defined as a function of string theory parameters using the DBI
action as follows
\begin{eqnarray}
g_{YM}^2=\frac{g_s}{\mu_4 \,(2\pi \alpha')^2\, \delta \tau},
\end{eqnarray}
where $\alpha'=l_s^2$ is the Regge slope parameter and $l_s$ is
the string length. Also, the 't Hooft coupling is
$\lambda=g_{YM}^2\, N_c$.

In AdS/QCD, there is gauge field living in the bulk AdS whose
dynamics is dual to the meson sector of QCD such as pions and higher
resonances. The gauge field on the $D4$ brane includes five
components, $A_\mu (\mu=0,1,2,3)$ and $A_U$. The $D4$ brane action
is given by
\begin{eqnarray}
S_{D4}&=& -\mu_4 \int d^5 x e^{-\phi} \sqrt{-\det(g_{MN}+2\pi\alpha'
F_{MN})}\nonumber\\ &&+S_{CS}, \label{FF}
\end{eqnarray}
where $F_{MN}=\partial_M A_N-\partial_N A_M-i[A_M,A_N],$
$(M,N=0,1,..5)$ is the field strength tensor, and the $A_M$ is the
$U(N_f)$ gauge field on the $D4$ brane. The second term in the above
action is the Chern-Simons action and $\mu_4=2\pi / (2\pi ls)^5$. It
is useful to define the new variable $z$ as
\begin{eqnarray}
U_z=(U_{KK}^5+U_{KK}^3\, z^2)^{1/5}.
\end{eqnarray}
Then by neglecting the higher order of $F^2$ in the expansion, the
$D4$ brane action can be written as
\begin{eqnarray}
S_{D4}= -\widetilde \mu_4(2\pi\alpha')^2
\int d^4x dz\,[\,
\frac{R^4}{4U_z^{5/2}}
\eta^{\mu\nu}\eta^{\rho\sigma} F_{\mu \rho}F_{\nu \sigma} \nonumber \\
+\frac{25}{8}\frac{U_z^{9/2}}{U_{kk}^3} \eta^{\mu\nu} F_{\mu
z}F_{\nu z}\,] +\mathcal{O}(F^3) \, ,
\end{eqnarray}
where $\widetilde \mu_4$ is
\begin{eqnarray}
\widetilde \mu_4=\sqrt{\frac{3}{2}} \, \frac{N_c U_{KK}^{3/2}}{5 R^3}\, \mu_4.
\end{eqnarray}
The gauge fields $A_\mu$ ($\mu=0,1,2,3$) and $A_z$ have a mode
expansion in terms of complete sets $\{\psi_n(z)\}$ and
$\{\phi_n(z)\}$ as
\begin{eqnarray}
A_\mu(x^\mu,z)&=&\sum_{n} B_\mu^{(n)}(x^\mu) \psi_n(z) \ ,
\label{expand;Av}\\
A_z(x^\mu,z)&=&\sum_{n} \varphi^{(n)}(x^\mu) \phi_n(z) \ .
\label{expand;A}
\end{eqnarray}
After calculating the field strengths, the action (10) is rewritten
as
\begin{eqnarray}
S_{D4}&=&-\widetilde \mu_4 (2\pi\alpha')^2 \int d^4x dz \sum_{m,n}
\bigg[ \frac{R^4}{4U_z^{5/2}}
F_{\mu\nu}^{(m)}F^{\mu\nu (n)}\psi_m\psi_n\nonumber \\
&&~~~
+\frac{25}{8}\frac{U_z^{9/2}}{U_{kk}^3}
(
\partial_\mu \varphi^{(m)}\partial^\mu \varphi^{(n)}\phi_m\phi_n \nonumber \\
&&~~~+B_\mu^{(m)}B^{\mu(n)}\dot\psi_m\dot\psi_n
-2\partial_\mu\varphi^{(m)}B^{\mu(n)}\phi_m\dot\psi_n )
\,\bigg],\nonumber\\
 \label{phiphiBB}
\end{eqnarray}
where the over dot denotes the derivative respect to the $z$
coordinate.

Let us consider first the vector meson field $B_\mu^{(m)}$. So, we
need to keep the following part of action:
\begin{eqnarray}
&&S_{D4} =-\widetilde \mu_4  (2\pi\alpha')^2 \int d^4x dz\,
\sum_{m,n} \nonumber\\
&& \times \bigg[\, \frac{R^4}{4U_z^{5/2}} F_{\mu\nu}^{(m)}F^{\mu\nu
(n)}\psi_m\psi_n  +\frac{25}{8}\frac{U_z^{9/2}}{U_{kk}^3}
B_\mu^{(m)}B^{\mu(n)}\dot\psi_m\dot\psi_n \bigg]. \nonumber\\
\label{Bonly}
\end{eqnarray}
We introduce the following dimensionless parameters:
\begin{equation}
\widetilde z \equiv\frac{z}{U_{KK}} \ ,
~~~ K(\widetilde z)\equiv 1+\widetilde z^2=\left(\frac{U_z}{U_{KK}}\right)^5 \ ,
\end{equation}
and using these parameters, we rewrite the action (15) as
\begin{eqnarray}
S_{D4}&=& -\widetilde \mu_4  (2\pi\alpha')^2\frac{R^4}{U_{KK}^{3/2}}
\int d^4x d\widetilde z \sum_{n,m}\, \nonumber\\
&& \bigg[
\frac{1}{4} K^{-1/2}F_{\mu\nu}^{(n)}F^{(m)\mu\nu}\psi_n\psi_m \nonumber\\
&&
+\frac{1}{2} M_{KK}^2 K^{9/10}  B_{\mu}^{(n)}B^{(m)\mu}
 \partial_{\widetilde z}\psi_n \partial_{\widetilde z}\psi_m \bigg] \ .
\end{eqnarray}
Functions $\psi_n$ ($n\ge 1$) satisfy the normalization condition as
\begin{equation}
\widetilde \mu_4 (2\pi\alpha')^2\frac{R^4}{U_{KK}^{3/2}} \int d\widetilde z\,  K^{-1/2}\, \psi_n\psi_m =\delta_{nm} \ .
\end{equation}
Also, we suppose the functions $\psi_n$ ($n\ge 1$) satisfy the
following condition
\begin{eqnarray}
\widetilde \mu_4 (2\pi\alpha')^2\frac{R^4}{U_{KK}^{3/2}} \int
d\widetilde z \,K^{9/10} \,\partial_{\widetilde z}
\psi_m\,\partial_{\widetilde z} \psi_n =\lambda_n\delta_{nm} \, .
\end{eqnarray}
Using Eqs. (18) and (19), an eigenvalue equation is obtained for the
functions $\psi_n$ ($n\ge 1$) as
\begin{equation}
-K^{1/2}\,\partial_{\widetilde z} \left(
K^{9/10}\,\partial_{\widetilde z} \psi_m\right)=\lambda_m\psi_m \ .
\end{equation}
Considering the above conditions, the action becomes canonically
normalized
\begin{eqnarray}
S_{D4}=
\int d^4x\, \sum_{n=1}^\infty
\left[\,
\frac{1}{4}
F_{\mu\nu}^{(n)}F^{\mu\nu (n)}
+\frac{1}{2}m_n^2\, B_\mu^{(n)}B^{\mu(n)}
\right]\, , \nonumber\\
\end{eqnarray}
where $B_\mu^{(n)}$ is a massive vector meson of mass $m_n\equiv
\lambda_n^{1/2} M_{KK}$ for all $n\ge 1$. Let us consider
$\varphi^{(n)}$ and rewrite the pseudo-scalar part of action (14) in
terms of new variables, Eq. (16):
\begin{eqnarray}
S_{D4}&=&-\widetilde \mu_4 (2\pi\alpha')^2
\int d^4x d\widetilde z \,\frac{25}{4}\, U_{KK}^{3/2} \, K^{9/10}  \nonumber\\
&& \times \sum_{m,n} \bigg[ \frac{1}{2} \, U_{KK} \partial_\mu
\varphi^{(m)}\partial^\mu \varphi^{(n)}\phi_m \phi_n
\nonumber\\
&&-\partial_\mu\varphi^{(m)} B^{\mu(n)}\phi_m
\partial_{\widetilde z}\psi_n
\bigg].\nonumber\\
\label{phiphiBB}
\end{eqnarray}
In order to normalize the kinetic part of the above action, we
consider the following orthonormal condition for $\phi_n$:
\begin{equation}
(\phi_m,\phi_n)\equiv \frac{25}{4}\,\widetilde \mu_4 (2\pi\alpha')^2
U_{KK}^{5/2} \int d\widetilde z \,K^{9/10}\,\phi_m \phi_n
=\delta_{mn} \,.
\end{equation}
By multiplying the Eq. (23) by $\lambda_n$ and comparing it with Eq.
(19), we find that the functions $\phi_{(n)}$ and $\dot\psi_n $ are
related together. In fact, we can consider
$\phi_n=m_n^{-1}\,\dot\psi_n$ ($n\ge 1$). Also, there exists a
function $\phi_0=C/K^{9/10}$ which is orthogonal to $\dot\psi_n$ for
all $n\ge 1$
\begin{equation}
(\phi_0,\phi_n)\propto \int d\widetilde z\, \partial_{\widetilde
z}\psi_n =0 \ , ~~~~(\mbox{for}~n\ge 1)\,  . \label{phi0}
\end{equation}
We use the normalization condition $1=(\phi_0,\phi_0)$ to obtain the
normalization constant $C$. Finally by using an appropriate gauge
transformation, the action (10) becomes
\begin{eqnarray}
S_{D4}&=&- \int d^4x \,
\bigg[\,
\frac{1}{2}\,\partial_\mu\varphi^{(0)}\partial^\mu\varphi^{(0)} \nonumber\\
&+& \sum_{n\ge 1}\bigg( \frac{1}{4} F_{\mu\nu}^{(n)}F^{\mu\nu (n)}
+\frac{1}{2} m_n^2\,B_\mu^{(n)}B^{\mu(n)} \bigg) \bigg], \label{S0}
\end{eqnarray}
where $\varphi^{(0)}$ is the pion field, which is the
Nambu-Goldstone boson associated with the chiral symmetry breaking.
An interpretation of this field is the same as the critical SS model
[39]. Therefore it is not necessary to repeat it here.

To ensure that the field strengths vanish at
$z\rightarrow\pm\infty$, it is useful to make another gauge choice,
namely the $A_z=0$ gauge. Actually, we can transform to the new
gauge through the following gauge transformation:
\begin{eqnarray}
A_M\rightarrow A_M -\partial_M \Lambda \ ,
\label{gaugetr}
\end{eqnarray}
and obtain the following new gauge fields:
\begin{eqnarray}
&&A_z(x^\mu,z)=0 \ ,\nonumber\\
&&A_\mu(x^\mu,z)=-\partial_\mu\varphi^{(0)}(x^\mu)\psi_0(z)
+\sum_{n\ge 1}
B_\mu^{(n)}(x^\mu)
\psi_n(z) \ .\nonumber\\
\label{expAA}
\end{eqnarray}
Function $\psi_0(z)$ is calculated through
\begin{equation}
\psi_0(z)=\int^z_0 dz'\, \phi_0(z') =C\,U_{KK}\,\widetilde z
\,F_1(0.5,0.9,1.5,-\widetilde z^2),
\end{equation}
where $F_1$ is  well-known hypergeometric function. It should be
noted that the massless pseudo scalar meson appears in the
asymptotic behavior of $A_\mu$, since we have
\begin{eqnarray}
A_\mu(x^\mu,z)\rightarrow \pm 1.8 C U_{KK} \,
\partial_\mu\varphi^{(0)}(x^\mu)~~(\mbox{as}~ z\rightarrow\pm\infty).
\label{Abdry}
\end{eqnarray}
In order to calculate the meson spectrum, it is necessary to solve
the Eq. (20) numerically by considering the normalization condition
(18).

Since Eq. (20) is invariant under $\widetilde z\rightarrow
-\widetilde z$, we can assume $\psi_n$ to be an even or odd
function. In fact, the $B_{\mu}^{(n)}$ is a four-dimensional vector
and axial vector if $\psi_n$ is an even or odd function,
respectively. The Eq. (20) is solved numerically using the shooting
method to obtain the mass of lightest mesons. Our results are
compared with the results of the SS model and experimental data in
Table I. As is clear, our result are in a good agreement with
experimental data. Also, the same values have been obtained in the
Ref. [49] using the $AdS_6$ background which is exactly coincident
with our results.

\begin{table}[htb]
\caption{\small . The ratio of the obtained eigenvalues of Eq. (20)
compared with the results of the SS model [39] and the ratio of
meson masses.} \center
\begin{tabular}{|c|c|c|c|}
\hline
$\,\,  \,\,$ &  $Our\,\, model$  &   $SS \,\,model$ &  $\,\, Experiment \,\,$
\\
\hline\hline
$\frac{\lambda_2}{\lambda_1}$ &  $  2.78   $ & 2.4  & $ \frac{m_{a_1(1260)}^2}{m_\rho^2}
\simeq\frac{(1230~ \mbox{MeV})^2}{(776~ \mbox{MeV})^2}
\simeq ~2.51  $ \\
\hline
$\frac{\lambda_3}{\lambda_1}$ & $  5.5 $   & 4.3  &  $\frac{m_{\rho(1450)}^2}{m_\rho^2}
\simeq\frac{(1465~ \mbox{MeV})^2}{(776~ \mbox{MeV})^2}
\simeq ~3.56 $ \\
\hline
$\frac{\lambda_3}{\lambda_2}$ & $  1.98 $   & 1.8  &  $\frac{m_{\rho(1450)}^2}{m_{a_1(1260)}^2}
\simeq\frac{(1465~ \mbox{MeV})^2}{(1230~ \mbox{MeV})^2}
\simeq ~1.41 $ \\
\hline
\end{tabular}
\end{table}

It is straightforward to generalize the above analysis to the case
of $N_f>1$ flavor QCD by introducing $N_f$ probe D4-branes. In order
to obtain a finite four-dimensional action for the modes localized
around $z=0$, the field strength $F_{MN}$ should vanish at
$z=\pm\infty$. This implies that the gauge field $A_M$ must
asymptotically take a pure gauge configuration
\begin{equation}
A_M(x^\mu,z)\rightarrow U_\pm^{-1}(x^\mu,z)\partial_M U_\pm(x^\mu,z)
\ , ~~~~~({\mbox{as}}~z\rightarrow\pm\infty).
\end{equation}
In analogy to the SS model [39], we can write
\begin{equation}
A_\mu(x^\mu,z)=U^{-1}(x^\mu)\partial_\mu U(x^\mu)\psi_+(z)
+\sum_{n\ge 1} B_\mu^{(n)}(x^\mu) \psi_n(z),\label{nonAexp}
\end{equation}
where
\begin{eqnarray}
\psi_{\pm}(z)=\frac{1}{2} \pm \widehat \psi_0(z),
\label{nonAexp}
\end{eqnarray}
\begin{eqnarray}
\widehat \psi_0(\widetilde z)=\frac{1}{3.6}\, \widetilde z \, F_1(0.5,0.9,1.5,-\widetilde z^2).
\label{nonAexp}
\end{eqnarray}
Now, by neglecting the vector meson fields, $B^{(n)}_\mu$ ($n\ge
1$), the field strengths can be written as
\begin{eqnarray}
F_{\mu\nu}&=&\left[U^{-1}\partial_\mu U,U^{-1}\partial_\nu U\right]\,
\psi_+(\psi_+-1) \ , \nonumber\\
F_{z \mu}&=&U^{-1}\partial_\mu U\,\partial_{\widetilde z} \widehat
\psi_0(\widetilde z).
\end{eqnarray}
Substituting these quantities into the non-Abelian generalization of
Eq. (10), we obtain
\begin{eqnarray}
S_{D4}= &-&\widetilde \mu_4 (2\pi\alpha')^2  \int d^4 x\,\mathop{\rm tr}\nolimits \bigg(
A(U^{-1}\partial_\mu U)^2\nonumber\\
&+&B\,[U^{-1}\partial_\mu U,U^{-1}\partial_\nu U\,]^2
\bigg) \ ,
\label{pionEA}
\end{eqnarray}
where the coefficients A and B are defined by the following
relations
\begin{eqnarray}
A&\equiv&2\, \frac{25}{8} \frac{1}{U_{KK}^3} \int d\widetilde z \,U_z^{9/2} ( \partial_{\widetilde z} \widehat \psi_0(\widetilde z))^2
=\frac{25 \, }{4 } \frac{U_{KK}^{1/2} }{3.6}\ ,\nonumber\\
B&\equiv&2\, \frac{R^4}{4} \int dz\,\frac{1}{U_z^{5/2}}\psi_+^2(\psi_+-1)^2
=\frac{0.16 R^4 }{2U_{KK}^{3/2}} \ .
\end{eqnarray}
If we compare the Eq. (35) with the familiar action of the Skyrme
model
\begin{eqnarray}
S&=&\int d^4 x\bigg(\frac{f_\pi^2}{4}\mathop{\rm tr}\nolimits (
 U^{-1}\partial_\mu U \,)^2\nonumber\\
 &+&\frac{1}{32 e^2}
\mathop{\rm tr}\nolimits  [U^{-1}\partial_\mu U,U^{-1}\partial_\nu
U]^2 \bigg) \ , \label{Skyrme}
\end{eqnarray}
it is possible to calculate the pion decay constant $f_\pi$ and
dimensionless parameter $e$ in terms of the non-critical model
parameters
\begin{eqnarray}
f_\pi^2= 4\,\widetilde \mu_4 (2\pi\alpha')^2 \,A=\sqrt{\frac32}\;
\frac{45\,\mu_4 (2\,\pi\,\alpha')^2}{3.6\,R^3} N_c M_{KK}^2\, ,
\label{fpi}
\end{eqnarray}
and
\begin{eqnarray}
\frac{1}{e^2}=32 \, \widetilde \mu_4 (2\pi\alpha')^2 \, B\, =
\sqrt{\frac38}\mu_4 (2\,\pi\,\alpha')^2\,R\, N_c\, . \label{e}
\end{eqnarray}
It is clear from the above equations that the parameters $f_\pi$ and
$e$ depend on $N_c$ as $f_\pi \sim \mathcal{O}(\sqrt{N_c})$ and
$e\sim\mathcal{O}(1/\sqrt{N_c})$ ,respectively. It is coincident
with the result obtained from the SS model and also QCD in large
$N_c$. We fix the $M_{KK}$ such that the $f_\pi\sim 93$ MeV for
$N_c=$3. So, we obtain $M_{KK}= 395$ MeV for our holographic model.
It should be noted that $M_{KK}$ is the only mass scale of the
non-critical model below which the theory is effectively pure
Yang-Mills in four dimensions.

\section{ Baryon in $AdS_6$ }

In this section we aim to introduce baryon configuration in the
non-critical holographic model. As is known, in the SS model the
baryon vertex is a $D4$ brane wrapped on a $S^4$ cycle. Here in
six-dimensional configuration, there is no compact $S^4$ sphere. So,
we introduce an unwrapped $D0$ brane as a baryon vertex instead
[26]. In analogy with the SS model, there is a Chern-Simons term on
the vertex world volume as
\begin{eqnarray}
S_{CS}\,\propto \, \int dt\, A_0(t),
\label{e}
\end{eqnarray}
which induces $N_c$ units of electric charge on the unwrapped $D0$
brane. In accordance with the Gauss constraint, the net charge
should be zero. So, one needs to attach $N_c$ fundamental strings to
the $D0$ brane. In turn, the other side of the strings should end up
on the probe $D4$ branes. The baryon vertex looks like an object
with $N_c$ electric charge with respect to the gauge field on the
$D4$ brane whose charge is the baryon number. This $D0$ brane
dissolves into the $D4$ brane and becomes an instanton soliton [26].
It is important to know the size of the instanton in our model. In
the SS model, it is shown that the size of an instantonic baryon
goes to zero at large 't Hooft coupling limit which is one of the
problems of the SS model in describing the baryons [23].

Let us consider the DBI action in the Yang-Mills approximation for
the $D4$ brane
\begin{equation}
S_{YM}=-\frac14\;\mu_4  (2\pi\alpha')^2 \; \int e^{-\phi}\,\sqrt{-g_{4+1}}  \;\rm tr \,F_{ mn}F^{ mn}\:.
\end{equation}
The induced metric on the $D4$ brane is
\begin{equation}
g_{4+1}=\left(\frac{U}{R}\right)^{2}\left(\eta_{\mu\nu}dx^{\mu}dx^{\nu}
+\left(\frac{R}{U}\right)^{4}\,\frac{dU^2}{f(U)}\right) \: .
\end{equation}
It is useful to define the new coordinate $w$
\begin{equation}
dw= \frac{R^{2}\,U^{1/2} \, dU }{\sqrt{U^5-U_{KK}^5}}\:.
\end{equation}
Using this coordinate, the metric (42) transforms to a conformally
flat metric
\begin{equation}
g_{4+1}=H(w)\left(dw^2+\eta_{\mu\nu}dx^{\mu}dx^{\nu}\right) \:,
\end{equation}
where
\begin{equation}
H(w)=(\frac{U}{R})^2 \:.
\end{equation}
Also, the $w$ coordinate can be rewritten in terms of the $z$
coordinate introduced in Eq. (16) as
\begin{equation}
dw=\frac25 \frac{R^{2}\,U_{KK}^{3} \, dz }{(U_{KK}^5-U_{KK}^3\, z^2)^{7/10}}\:.
\end{equation}
Note that in the new conformally flat metric, the fifth direction is a finite interval $[-w_{max}, w_{max}]$ because
\begin{eqnarray}
w_{max}&=&\int_0^\infty \frac{R^{2}\,U^{1/2} \, dU }{\sqrt{U^5-U_{KK}^5}} \nonumber \\
&=&\frac{R^{2}}{U_{KK} } \int_1^\infty \frac{d\tilde U}{\sqrt{\tilde U^5-1}}
\simeq \frac{R^{2}}{U_{KK}}1.25 <\infty .
\end{eqnarray}
We can approximate $w$ near the origin $w\simeq 0$, as
\begin{equation}
w\simeq \frac{2}{5}\left(\frac{R}{U_{KK}}\right)^{2} z,
\end{equation}
and using relation (6), we obtain
\begin{equation}
w\simeq \frac{z}{M_{KK} \, U_{KK}} \,\,\,\,\,or\,\,\,\,\,M_{KK}\, w\simeq \frac{z}{ U_{KK}},
\end{equation}
or equivalently,
\begin{equation}
U^5\simeq U_{KK}^5(1+M_{KK}^2\,w^2)\, .
\end{equation}
In analogy with the SS model, this relation implies that $M_{KK}$ is
the only mass scale that dictated the deviation of the metric from
the flat configuration and it is the only mass scale of the theory
in the low energy limit.(It should be noted that the $D4$ branes
come with two asymptotic regions at $w\rightarrow \pm w_{max}$
corresponding to the ultraviolet and infrared region near the $w
\simeq 0$.)

Equation (41) is rewritten in the conformally flat metric (44) as
\begin{eqnarray}
S_{YM}^{D4}&&=-\frac14\mu_4  (2\pi\alpha')^2  \int d^4x dw e^{-\phi} \left(\frac{U(w)}{R}\right)\rm tr F_{mn}F^{mn} \nonumber \\
&&=-\;\int dx^4 dw \;\frac{1}{4e^2(w)} \;\rm tr F_{mn}F^{mn} \: .
\end{eqnarray}
Thus, the position dependent electric coupling $e(w)$ of this five
dimensional Yang-Mills is equal to
\begin{equation}
\frac{1}{e^2(w)}\equiv \frac{ \sqrt{3/2} \; \mu_4\,(2\,\pi\,\alpha')^2\,R\,N_c}{5 }\; M_{KK} \left( \frac{U}{ U_{KK}}\right) \:.
\end{equation}
Also, for a unit instanton we have
\begin{equation}
\frac{1}{8\pi^2 }\int \rm tr F\wedge F=\frac{1}{16\pi^2} \int \rm tr
F_{mn} F^{mn} =1 .
\end{equation}
Inserting the above relations in the Eq. (51), we obtain the energy
of a point-like instanton localized at $w=0$ as
\begin{equation}
m_B^{(0)}= \frac{ \sqrt{3/2} \;4\pi^2 \mu_4\,(2\,\pi\,\alpha')^2 R }{5}\;N_c  \; M_{KK}\:.
\end{equation}
By increasing the size of the instanton, more energy is needed
because $1/e^2(w)$ is an increasing function of $|w|$. So the
instanton tends to collapse to a point-like object. On the other
hand, $N_c$ fundamental strings attached to the $D4$ branes behave
as $N_c$ units of electric charge on the brane. The Coulomb
repulsions among them prefer a finite size for the instanton.
Therefore, there is a competition between the mass of the instanton
and Coulomb energy of fundamental strings. For a small instanton of
size $\rho$ with the density $D(x^i,w)\sim
\rho^4/(r^2+w^2+\rho^2)^4$, the Yang-Mills energy is approximated as
\begin{equation}\label{mass}
\sim \frac16\, m_B^{(0)}M_{KK}^2\rho^2\:,
\end{equation}
and the five dimensional Coulomb energy is
\begin{equation}\label{C}
\sim \frac12\times \frac{e(0)^2N_c^2}{10\pi^2\rho^2}\:.
\end{equation}
The size of a stable instanton is obtained by minimizing the total
energy
\begin{equation}\label{size}
\rho^2_{baryon}\simeq \frac{1}{\sqrt{3/2} \;2\pi^2
\mu_4\,(2\,\pi\,\alpha')^2} \frac{1}{M_{KK}^2}\, .
\end{equation}
As it is stated in the previous section, in the SS model (the
critical version of dual QCD) the size of the instanton goes to zero
because of the large 't Hooft coupling limit. However in the
non-critical string theory, the 't Hooft coupling is of order one.
So, the size of the instanton is also of order 1 but it is still
smaller than the effective length of the fifth direction $\sim 1/
M_{KK}$ of the dual QCD.

\section{ Nucleon-Nucleon potential }

In the previous section, we demonstrated that the size of the baryon
in the non-critical holographic model is smaller than the scale of
the dual QCD and we can assume that the baryon is a point-like
object in five dimensions. Thus as a leading approximation, we can
treat it as a point-like quantum field in five dimensions. In the
rest of this paper, we will restrict ourselves to fermionic baryons
because we intend to study the nucleons. So, we consider the odd
$N_c$ to study a fermionic spin $1/2$ baryon. We choose $N_c=3$ in
our numerical calculations for realistic QCD. Also, we will assume
$N_F=2$ and consider the lowest baryons which form the
proton-neutron doublet under $SU(N_F=2)$. All of these assumptions
lead us to introduce an isospin $1/2$ Dirac field, $\cal N$ for the
five dimensional baryon.

The leading 5D kinetic term for $\cal N$ is the standard Dirac
action in the curved background along with a position dependent mass
term for the baryon. Moreover, there is a coupling between the
baryon field and the gauge filed living on the flavor branes that
should be considered. Therefore, a complete action for the baryon
reads as
\begin{eqnarray}
\int d^4 x dw\bigg[&-&i\bar{\cal N}\gamma^m D_m {\cal N}
-i m_b(w)\bar{\cal N}{\cal N} \nonumber \\
&+&g_5(w){\rho_{baryon}^2\over
e^2(w)}\bar{\cal N}\gamma^{mn}F_{mn}{\cal N} \bigg] \nonumber \\
&-&\int d^4 x  dw {1\over 4 e^2(w)} \;\rm tr\, F_{mn}F^{mn}\,,
\label{5dfermion1}
\end{eqnarray}
where $D_m$ is a covariant derivative, $\rho_{baryon}$ is the size
of the stable instanton, and $g_5(w)$ is an unknown function with a
value at $w=0$ of  $2 \pi^2 /3$ [23]. $\gamma^m$ are the standard
$\gamma$  matrices in the flat space and $\gamma^{mn}=1/2
[\gamma^m,\gamma^n]$.

The factor $\rho_{baryon}^2\over e^2(w)$ is used for convenience.
Usually, the first two terms in the action are called the minimal
coupling and the last term in the first integral refers to the
magnetic coupling.

A four dimensional nucleon is the localized mode at $w\simeq 0$
which is the lowest eigenmode of a five dimensional baryon along the
$w$ direction. So, the action of the five dimensional baryon must be
reduced to the Four dimension. In order to do this, one should
perform the KK mode expansion for the baryon field $\mathcal
N(x_{\mu},w)$ and the gauge field $A(x_{\mu},w)$. The gauge field
has a KK mode expansion which studied in Sec. (3) in detail. The
baryon field also can be expanded as
\begin{eqnarray}
{\cal N}_{L,R}(x^\mu,w)=N_{L,R}(x^\mu)f_{L,R}(w),
\label{5dfermion1}
\end{eqnarray}
where $N_{L,R}(x^\mu)$ is the chiral component of the four
dimensional nucleon field. Also the profile functions, $f_{L,R}(w)$
satisfy the following conditions:
\begin{eqnarray}
\partial_w f_L(w)+m_b(w) f_L(w) &=& m_B f_R(w)\:,\nonumber\\
-\partial_w f_R(w)+m_b(w) f_R(w) &=& m_B f_L(w)\:,
\end{eqnarray}
in the range $w\in[-w_{max},w_{max}]$, and the eigenvalue $m_B$ is
the mass of the nucleon mode, $N(x)$. Moreover, the eigenfunctions
$f_{L,R}(w)$ obey the following normalization condition
\begin{equation}
\int_{-w_{max}}^{w_{max}} dw\,\left|f_L(w)\right|^2 =
\int_{-w_{max}}^{w_{max}} dw\,\left|f_R(w)\right|^2 =1\:.
\end{equation}
It is more useful to consider the following second-order
differential equations for $f_{L,R}(w)$
\begin{eqnarray}
&&\left[-\partial^2_w -\partial_w m_b(w)+(m_b(w))^2\right]
f_L(w)=m_B^2 f_L(w)\:,\nonumber \\
&&\left[-\partial^2_w +\partial_w m_b(w)+(m_b(w))^2\right]
f_R(w)=m_B^2 f_R(w)\:.\nonumber\\
\label{eigeneq}
\end{eqnarray}

As we approach $w\to \pm w_{max}$, $m_b(w)$ diverges as $m_b(w) \sim
{1\over (w\mp w_{max})^2}$ and the above equations have normalizable
eigenfunctions with a discrete spectrum of $m_B$. Note that the term
$-\partial_w m_b(w)$ is asymmetric under $w\to -w$. It causes that
$f_L(w)$  tends to shift to the positive side of $w$ and the
opposite behavior happens for $f_R(w)$. It is important in the axial
coupling of the nucleon to the pions.

It is mentioned in Sec. (2) that the gauge field has a mode
expansion (31) at $A_z=0$ gauge which can be rewritten as
\begin{equation}
A_\mu(x,w)=i\alpha_\mu(x)\psi_0(w) +i\beta_\mu(x)+\sum_n
B_\mu^{(n)}(x)\psi_{(n)}(w)\:,
\end{equation}
where $\alpha_\mu$ and $\beta_\mu$ are related to the pion field
 $U(x)=e^{2i\pi(x)/f_\pi}$ by the following relations,
\begin{eqnarray}
&&\alpha_\mu(x)\equiv \{ U^{-1/2},\partial_{\mu} U^{1/2}\}\, ,\nonumber\\
&&\beta_\mu(x)\equiv \,\frac 12 [ U^{-1/2},\partial_{\mu}
U^{1/2}]\, .
\end{eqnarray}

Here, we use the above expansion along with the properties of
 $f_L(w)=\pm f_R(-w)$, $\psi_0$ and $\psi_{(n)}$ under the
$w\to-w$ transformation to calculate the four dimensional action. It
is worthwhile to note that again $\psi_{(2k+1)}(w)$ is even, while
$\psi_{(2k)}(w)$ is odd under $w\to-w$, corresponding to vector
$B_\mu^{(2k+1)}(x^\mu)$ and axial-vector mesons
$B_\mu^{(2k)}(x^\mu)$ respectively. For simplicity, we neglect the
Chern-Simons term in the baryon action, Eq. (58). By inserting the
mode expansion of the baryon field in the action, we obtain the
minimal coupling as
\begin{eqnarray}
&&S^{min}\nonumber\\
&&~~ = \int d^4 x dw  \left[-i\bar{ N} \bar f \gamma^m
(\partial_m-iA_m) { N} f
-i m_b(w)\bar{ N }\bar f { N}f \right]\nonumber\\
&&~~~~~~~~~~~~=\int d^4 x \left[-i\bar{ N}\, \gamma^{\mu} \partial_{\mu} { N}-i m_B \bar{ N } \, { N} \right]\nonumber\\
&&~~~~~~~~~~~~~~~~~~~-\int d^4 x dw \left[\bar{ N}\,\bar f\,
\gamma^{\mu} A_{\mu} { N}\,f \right] \, .\label{5dfermion1}
\end{eqnarray}
Now, we expand the gauge field presented in the last integral using
the Eq. (63). Since the parity of $\psi_{(n)}(w)$ depends on $n$, it
is possible to separate the odd and even $n$. After taking the
integrals over $w$, we obtain the four dimensional minimal action
for the nucleon as
\begin{eqnarray}
S^{min}= \int d^4x \bigg[ &-&i\bar N \gamma^\mu \partial_\mu N- im_B \bar N N\nonumber\\
&+&{\cal L}_{\rm vector}^{min}+{\cal L}_{\rm axial}^{min} \bigg] \:,
\end{eqnarray}
where the minimal vector and axial interactions are
\begin{eqnarray}
{\cal L}_{\rm vector}^{min}=&-&i\bar N \gamma^\mu \beta_\mu
 N-\sum_{k\ge 0}g_{V,min}^{(k)} \bar N \gamma^\mu  B_\mu^{(2k+1)} \,N \, ,\nonumber\\
{\cal L}_{\rm axial}^{min}=&-&\frac{i g_{A,min}}{2}\bar N
\gamma^\mu\gamma^5 \alpha_\mu N
\nonumber\\
&-&\sum_{k\ge 1} g_{A,min}^{(k)} \bar N \gamma^\mu \, \gamma^5\,
B_\mu^{(2k)} N \, . \label{vector-axial-coupling}
\end{eqnarray}
The various minimal couplings constants
$g_{V,min}^{(k)},g_{A,min}^{(k)}$ as well as the pion-nucleon axial
coupling $g_{A,min}$ are calculated by the following suitable
overlap integrals of wave functions
\begin{eqnarray}
g_{V,min}^{(k)}&=&\int_{-w_{max}}^{w_{max}} dw\,\left|f_L(w)\right|^2
\psi_{(2k+1)}(w)\:,\nonumber\\
g_{A,min}^{(k)}&=&\int_{-w_{max}}^{w_{max}} dw\,\left|f_L(w)\right|^2
\psi_{(2k)}(w)\:,\nonumber\\
g_{A,min}&=&2\int_{-w_{max}}^{w_{max}} dw\,\left|f_L(w)\right|^2
\psi_0(w)\:.
\end{eqnarray}
Also, the magnetic interaction term in Eq. (58) becomes
\begin{equation}
S^{magnetic}= - \int d^4 x dw\left( g_5(w){\rho_{baryon}^2\over
e^2(w)} \bar N \bar f \gamma^{w \mu}F_{w \mu} N f \right).
\label{5dfermion1}
\end{equation}
Inserting the gauge field expansion into Eq. (69), the magnetic
interaction reads as
\begin{equation}
S^{magnetic}= \int d^4x \left({\cal L}_{\rm vector}^{magnetic}+{\cal L}_{\rm axial}^{magnetic} \right) \:,
\end{equation}
where
\begin{eqnarray}
{\cal L}_{\rm vector}^{magnetic}&=& -\sum_{k\ge 0}g_{V,mag}^{(k)} \bar N \gamma^\mu \gamma^5 B_\mu^{(2k+1)} \,N \,,\nonumber\\
{\cal L}_{\rm axial}^{magnetic}&=&-\frac{i g_{A,mag}}{2} \bar N  \gamma^\mu \gamma^5 \alpha_\mu N \nonumber\\
&-&\sum_{k\ge 1} g_{A,mag}^{(k)} \bar N \gamma^\mu\gamma^5
B_\mu^{(2k)} N \, , \label{vector-axial-coupling}
\end{eqnarray}
and the magnetic couplings are defined as
\begin{eqnarray}
g_{A,mag}&=& 4 \int_{-w_{max}}^{w_{max}}
 dw \left(g_5(w){\rho_{baryon}^2\over
e^2(w)} \right) \nonumber\\ && ~~~~~~~~~~~~~~~~~~
\times \left|f_L(w)\right|^2\partial_w \psi_0(w) ,\nonumber\\
 \nonumber\\
 g_{A,mag}^{(k)}&=& 2 \int_{-w_{max}}^{w_{max}}
 dw \left(g_5(w){\rho_{baryon}^2\over
e^2(w)} \right)  \times  \nonumber\\ && ~~~~~~~~~~~~~~~~~~ \times
\left|f_L(w)\right|^2\partial_w \psi_{(2k)}(w)\, ,\nonumber\\
\nonumber\\ g_{V,mag}^{(k)}&=& 2 \int_{-w_{max}}^{w_{max}}
 dw \left(g_5(w){\rho_{baryon}^2\over
e^2(w)} \right) \times  \nonumber\\ && ~~~~~~~~~~~~~~~~~~ \times
\left|f_L(w)\right|^2\partial_w \psi_{(2k+1)}(w)\, .
\end{eqnarray}
Using Eq. (52), we can rewrite the magnetic couplings as
\begin{eqnarray}
g_{V,mag}^{(k)}&=& 2\,C_{mag} \int_{-w_{max}}^{w_{max}}
dw \left( \frac{g_5(w)}{g_5(0)}\right)\left( \frac{U(w)}{U_{KK}}\right) \nonumber\\
&& ~~~~~~~~~~~~~~~~~~~~~~ \times  \left|f_L(w)\right|^2\partial_w
\psi_{(2k+1)}(w)\:,
\nonumber\\ \nonumber\\
g_{A,mag}^{(k)}&=&
2\,C_{mag} \int_{-w_{max}}^{w_{max}}
dw \left( \frac{g_5(w)}{g_5(0)}\right)\left( \frac{U(w)}{U_{KK}}\right)  \nonumber\\
&& ~~~~~~~~~~~~~~~~~~~~~~ \times \left|f_L(w)\right|^2\partial_w
\psi_{(2k)}(w)\:,
\nonumber\\ \nonumber\\
g_{A,mag}&=& 4\, C_{mag}\int_{-w_{max}}^{w_{max}}  dw \left(
\frac{g_5(w)}{g_5(0)}\right)\left( \frac{U(w)}{U_{KK}}\right)
\nonumber\\ && ~~~~~~~~~~~~~~~~~~~~~~\times
\left|f_L(w)\right|^2\partial_w \psi_0(w)\,,
\end{eqnarray}
where we define $C_{mag}$ as
\begin{eqnarray}
C_{mag}=\frac{ \sqrt{3/2} \mu_4\,(2\,\pi\,\alpha')^2}{5 }\,R\,N_c\;g_5(0)\, M_{KK} \;\rho_{baryon}^2\:.
\end{eqnarray}
Also, there is a next-to-leading order term in the magnetic coupling
action which is responsible for the derivative couplings. Finally by
considering the derivative terms, the Lagrangian of the nucleon is
obtained as
\begin{eqnarray}
{\cal L}_{\rm Nucleon}= -i\bar N \gamma^\mu \partial_\mu N- im_B \bar N N+
{\cal L}_{\rm vector}+{\cal L}_{\rm axial}\,,\nonumber\\
\end{eqnarray}
where
\begin{eqnarray}
{\cal L}_{\rm vector}=&-&i\bar N \gamma^\mu \beta_\mu
 N-\sum_{k\ge 0}g_{V}^{(k)} \bar N \gamma^\mu  B_\mu^{(2k+1)} \,N \nonumber\\
 &+&\sum_{k\ge 0}g_{dV}^{(k)} \bar N \gamma^{\mu \nu} \partial_\mu B_\nu^{(2k+1)}
 \,N\, ,
 \nonumber\\
{\cal L}_{\rm axial}&=&-\frac{i g_{A}}{2}\bar N  \gamma^\mu\gamma^5 \alpha_\mu N
-\sum_{k\ge 1} g_{A}^{(k)} \bar N \gamma^\mu\gamma^5 B_\mu^{(2k)} N \nonumber\\
&+&\sum_{k\ge 0}g_{dA}^{(k)} \bar N \gamma^{\mu \nu} \gamma^5
\partial_\mu B_\nu^{(2k)} \,N \, .\label{vector-axial-coupling}
\end{eqnarray}
Also, $g=g_{min}+g_{mag}$ stands for all the couplings. We neglect
the derivative couplings in the following calculations as a leading
approximation.

Since the instanton carries only the non-Abelian field strength, the
iso-scalar mesons couple to the nucleon in a different formalism
than the iso-vector mesons. Therefore for the iso-scalar mesons,
such as the $\omega^{(k)}$ meson, only the minimal couplings
contribute
\begin{eqnarray}
g_{A}^{iso-scalar}&=&g_{A,min}\:,\nonumber\\
g_{A}^{(k),iso-scalar}&=&g_{A,min}^{(k)}\:,\nonumber\\
g_{V}^{(k),iso-scalar}&=&g_{V,min}^{(k)}\: .
\end{eqnarray}
However, the iso-vector mesons couple to the nucleon from both the
minimal and magnetic channels. Thus, iso-vector meson couplings are
\begin{eqnarray}
g_{A}^{iso-vector}&=&g_{A,min}+g_{A,mag}\:,\nonumber\\
g_{A}^{(k),iso-vector}&=&g_{A,min}^{(k)}+g_{A,mag}^{(k)}\:,\nonumber\\
g_{V}^{(k),iso-vector}&=&g_{V,min}^{(k)}+g_{V,mag}^{(k)}\: .
\end{eqnarray}

The iso-scalar and iso-vector mesons have the same origin in the
five dimensional dynamics of the gauge field. In fact, if we write
the gauge field in the fundamental representation, we could
decompose the massive vector mesons as
\begin{equation}
B_\mu^{(2k+1)}=\left(\begin{array}{cc}1/2 &0 \\
0&1/2\end{array}\right) \omega^{(k)}_\mu+\rho_\mu^{(k)}\, ,
\end{equation}
where $\omega^{(k)}_\mu$ and $\rho_\mu^{(k)}$ are the iso-scalar and
the iso-vector parts of a vector meson, respectively. Since the
baryon is made out of $N_c$ product quark doublets, the above
composition for nucleon should be written as
\begin{equation}
B_\mu^{(2k+1)}=\left(\begin{array}{cc}N_c/2 &0 \\ 0&N_c/2\end{array}\right)
\omega^{(k)}_\mu+\rho_\mu^{(k)}\:.
\end{equation}
Therefore, there is an overall factor $N_c$ between the
iso-scalar, $\omega^{(k)}_\mu$ and iso-vector, $\rho_\mu^{(k)}$
mesons. Indeed, there is a universal relation between the Yukawa
couplings involving the iso-scalar and iso-vector mesons
\begin{equation}
|g_{\omega^{(k)}NN}|\simeq  N_c \times |g_{\rho^{(k)}NN}|\,
.\label{prediction-gomega}
\end{equation}
We will be back to this relation later. Here we need to solve the
eigenvalue equation (62) numerically to obtain the wave function,
$f_{L,R}$ and the mass, $m_B$ of the nucleon. It is useful to define
the dimensionless variables $\tilde w=M_{KK}w$, $\tilde
U={U/U_{KK}}$, and $\tilde z={z/ U_{KK}}$ which are related together
by
\begin{equation}
\tilde w = \int_0^{\tilde z}\, {d\tilde z \over \left[1+\tilde
z^2\right]^{7\over 10}} = {5\over 2}\int_1^{\tilde U}\,d\tilde U
\sqrt{\frac{\tilde U}{\tilde U^5 -1}}\:,
\end{equation}
and rewrite the $m_b(w)$ in terms of these dimensionless variables
as
\begin{equation}
m_b(w)\simeq m_b^{(0)}\cdot\tilde U=M_{KK}\; \tilde m_b(\tilde w),
\end{equation}
where
\begin{equation}
\tilde m_b(\tilde w)=\frac{ \sqrt{3/2} \;4\pi^2 \mu_4\,(2\,\pi\,\alpha')^2\,R}{5 }\;\,N_c \; \tilde U(\tilde w).
\end{equation}
After rewriting Eq. (62) in terms of $\tilde w$, we obtain
\begin{equation}
\left[-\partial^2_{\tilde w} -\, \partial_{\tilde w}\tilde
m_b({\tilde w})+(\tilde m_b({\tilde w}))^2\right] f_L(\tilde
w)=\left(m_B\over M_{KK}\right)^2 f_L(\tilde w)\:.\label{redeigeneq}
\end{equation}
The key idea for using dimensionless variables is that the function
$f_L(\tilde w)$ does not depend on the scales further. Now, we use
the shooting method again to solve the above equation numerically
and find $f_L(\tilde w)$ and its eigenvalue, ${m_B/ M_{KK}}$. In
order to do the numerical calculation, we assume $N_c=3$ for
realistic QCD. Also as was mentioned in the previous section, we
choose the value of $M_{KK}=0.395$ GeV to have the pion decay
constant $f_{\pi}=0.093$ GeV. We obtain the various couplings by
evaluating integrals (68) and (73) and compare some of our results
with the results of the SS model [23] in Table II.

\begin{table}[htb]
\caption{\small . Numerical results for axial and vector meson
couplings in the non-critical holographic model of QCD. The values
of vector couplings are compared with the SS model results[23]. }
\center
\begin{tabular}{|c|c|c|c|c|c|c|}
\hline $k$  &$g^{(k)}_{A,min}$ & $g_{A,mag}^{(k)}$ & $g^{(k),a}_{V,min}$ & $g^{(k),b}_{V,min}$ & $g_{V,mag}^{(k),a}$ & $g_{V,mag}^{(k),b}$ \\
 \hline\hline
0& 1.16 & 1.86 &8.30 & 5.933 & -1.988 & -0.816  \\
 \hline
1 &  1.07& 1.44 & 1.6488 &3.224& -6.83 & -1.988 \\
 \hline
2 &  0.96  & 0.862& 1.9 & 1.261& -7.44 &-1.932 \\
\hline
3  & 0.67 & 0.14& 0.688 & 0.311 & -4.60 & -0.969 \\
\hline
\end{tabular}
\\
(a) presented model results \\
(b) SS  model results
\end{table}

Also, using this non-critical model, the axial couplings are
obtained as
\begin{equation}
g_{A,mag}=1.582 \,\,\,,\,\,\, g_{A,min}\simeq 0 \, ,
\end{equation}
while in the previous analysis [23] using the SS model, these
couplings are reported as
\begin{equation}
g_{A,mag}=0.7\, \frac{N_c}{3} \,\,\,,\,\,\, g_{A,min}\simeq 0.13
\, .
\end{equation}
If we choose $N_c=3$, then the SS model predicts $g_{A,mag}=0.7$ and
$g_{A}=0.83$. It should be noted  that the higher order of $1/{N_c}$
corrections can be used to improve this result but the lattice
calculations indicate that higher order of $1/{N_c}$ corrections are
suppressed. Our results are a good approximation of the experimental
data at leading order $g_{A}^{exp}=1.2670 \pm 0.0035$.

\section{ Nucleon-meson couplings }

In the previous section, we explained that the NN interaction can be
interpreted as meson exchange potentials. We showed that the
nucleons couple to the meson through the minimal and magnetic
couplings. Our holographic NN potential contains just the vector,
axial-vector, and pseudo-scalar meson exchange potentials which have
the isospin dependent and isospin independent components.

All of the leading order meson-nucleon couplings are calculated
numerically and compared with the predictions of the four modern
phenomenological NN interaction models such as the AV 18 [8],
CD-Bonn [7], Nijmegen(93) [6] and Paris [5] potentials in Table III.
Also, results of the SS model are presented in the table. It is
necessary to mention here that the components of the
phenomenological models are very different in strength, and if
parameterized in terms of single meson exchange give rise to
effective meson-nucleon coupling strengths, which also are similar.
We explain different components of the NN potential below in detail.

\begin{table}[htb]
\caption{\small . The values of different effective meson-nucleon
couplings in the phenomenological interaction models [54], SS model
[23], and our model. } \center
\begin{tabular}{|c|c|c|c|c|c|c|}
\hline g &$V\,18$ & $CD-Bonn$ &  $Nijm\,(93)$  & $Paris$ & $SS\, model$ & $Our\, model$ \\
 \hline\hline
$g_{a^0} $ & 9.0 & 9.0 & 9.0 & 10.4 & - & - \\ \hline $g_\sigma$
&9.0 & 11.2 & 9.8 & 7.6 & - & - \\ \hline $g_\pi$    & 13.4 &
13.0&12.7&13.2 &  16.48   & 15.7  \\ \hline $g_\eta$   &  8.7 & 0.0
& 1.8 &11.7& 16.13 & 0.0 \\ \hline $g_\omega$ & 12.2 & 13.5& 11.7
&12.7 & 12.6 & 11.57 \\ \hline $g_\rho$   & -    &3.19 & 2.97
&- & 3.6 & 3.15 \\ \hline $g_{a^1}$& - & - &- &- & 3.94 & 1.51 \\
\hline
$g_{f^1}$  & - & - & - & - &  & 1.74 \\
\hline
\end{tabular}
\end{table}

\subsection{ Scalar potential}

In the phenomenological interaction models, exchange of a single
scalar meson produces the isospin independent scalar potential. The
mass of the lowest scalar meson is not established well [55], but in
the phenomenological NN interaction models it is typically taken to
be of order $500-700$ MeV. By considering $m_S$ to be roughly 600
MeV, the effective scalar meson coupling constant for these
interaction models differs from 7.6 (Paris) to 11.2 (CD-Bonn). It
should be noted again that these values are the effective couplings.
In fact two of these models do not contain any scalar meson in their
parameterized forms. In our holographic model based on the
non-critical string theory, there is no scalar interaction term
either.

The isospin dependent component of the scalar potential can be
interpreted as a scalar spin 1 meson exchange. In $N_c$ counting,
this component is of order $1/N_c$, so it is weaker than the isospin
independent scalar meson by an order of its magnitude. The effective
values for the lowest scalar meson [$a_0(980)$], range from 9.0 to
10.4 in the various phenomenological NN interaction models.

\subsection{ Vector potential}

The vector component of the phenomenological NN interaction models
is equal to the scalar component with the minus sign. It means that
the strength of the $a_0(980)$ exchange interaction is equal to the
exchange of $\rho (770)$ in the large $N_c$ limit. Indeed, the
vector potential arises from a stronger $\omega$ meson exchange
(isospin independent component) and a weaker $\rho$ meson exchange
(isospin dependent component) interaction. In our model the vector
meson couplings are related to the minimal and magnetic couplings as
follows
\begin{eqnarray}
g_{\omega^{(k)}NN}&\equiv & \frac{N_c\; g_{V}^{(k),iso-scalar}}{2} =\,\frac{N_c\; g_{V,min}^{(k)}}{2} \:,\nonumber\\
g_{\rho^{(k)}NN}&\equiv  & \frac{ g_{V}^{(k),iso-vector}}{2}
=\,\frac{ g_{V,min}^{(k)}+g_{V,mag}^{(k)}}{2} \: .
\end{eqnarray}

The value of effective $\omega-$ nucleon coupling ranges from 11.7
[Nijmegen(93)] to 13.5 (CD-Bonn), while in the original version of
these models this value varies from 10.35 (Nijmegen) to 15.85
(Bonn)[56]. In the SS model, $g_\omega$ is equal to 12.6 by
considering $M_{KK}=940$ MeV, $N_c=3$, and $\lambda=17$. We also
obtain the value of $g_\omega=$ 11.57 which is in the range of the
values anticipated from the phenomenological potential models. We
have used $N_c=3$ and $M_{KK}=395$ MeV in our calculations. The
obtained value for the nucleon mass in our model is roughly 920 MeV
which is very realistic and close to the familiar nucleon mass.

The isospin dependent component of the vector potential which arises
from a $\rho$ meson exchange is roughly three times weaker than the
isospin independent component. In a chiral quark model, it is
expected to have $g_\omega=3\, g_\rho$, but the value of the
$\mathcal{R}=g_\omega/3\, g_\rho$ differs from the one in the above
phenomenological interaction models. It is 1.66 for the CD-Bonn, 1.5
for the Nijmegen, and 0.77 in the Paris model. This ratio is about
1.2 in the SS model and equals to $\mathcal{R}=1.33$ in our model.
Actually, the NN phase shifts uniformly require a larger
$\mathcal{R}$ than the chiral quark model prediction which is a
mystery. However in the resultant potential of the holographic QCD
model, it can be explained by the contribution of the magnetic
coupling in the vector channel.

\subsection{ Axial-vector potential}

The Nijmegen(93) and CD-Bonn models do not contain any single axial
vector meson exchange, so there is no axial vector interaction in
their structures. The phenomenological AV 18 and Paris potentials
predict a small value for the axial vector interactions too.

In our holographic potential model, the axial vector mesons
$a^{(k)}$ and $f^{(k)}$ couple to the nucleon with the following
couplings:
\begin{eqnarray}
g_{f^{(k)}NN}&\equiv & \frac{N_c \, g_{A}^{(k),iso-scalar}}{2}
=\,\frac{N_c\, g_{A,min}^{(k)}}{2}, \nonumber\\
g_{a^{(k)}NN}&\equiv & \frac{g_{A}^{(k),iso-vector}}{2} =\,\frac{
g_{A,min}^{(k)}+g_{A,mag}^{(k)}}{2}\:.
\end{eqnarray}

The values of the $a_{(1)}-$ nucleon, and $f_{(1)}-$ nucleon
couplings are obtained at about 1.51 and 1.74, respectively, in our
holographic potential.

\subsection{ Pseudo-scalar potential}

The isospin independent pseudo-scalar interaction comes from an
$\eta'$ meson exchange. This component is not well constrained by NN
scattering data and the phenomenological interaction models give
extremely different values for this component ranging from 1.8
[Nijmegen(93)] to 11.7 (Paris). While analysis of other observables
such as $\eta'-$meson photo-production suggest that the coupling
constant value should not go beyond the 2.2 [57]. In our holographic
potential, the pseudo-scalar mesons such as the pion $\pi$ and
$\eta'$ couple to the nucleon as
\begin{eqnarray}
\frac{g_{\pi^{(k)}NN}}{2 m_N} M_{KK}  &\equiv & \frac{ g_{A}^{iso-vector}}{2 f_{\pi}} M_{KK} = \frac{ g_{Amin}+g_{A,mag} }{2  f_{\pi}}  M_{KK} ,\nonumber\\
\frac{g_{\eta'^{(k)}NN}}{2 m_N} \,M_{KK} &\equiv  & \frac{N_c\, g_{A}^{iso-scalar}}{2\, f_{\pi}}\,M_{KK} =\,\frac{N_c g_{A,min}}{2\, f_{\pi}}\,M_{KK} .\nonumber\\
\end{eqnarray}

We obtain $g_\eta=0$ at the leading order. Despite the isospin
independent component, the isospin dependent component is strong and
spread on a long range. All of the models considered here have a
main component for this interaction which is a pion exchange
potential. The values of the pion-nucleon coupling constant, $g_\pi$
vary from the 12.7 to 13.4 effectively. In our calculation, $g_\pi$
is evaluated as 15.7 while in the calculations based on the SS model
it is obtained at 16.48 [23].

\section{ Conclusion }

In this study, we used the non-critical holographic duality of gauge
theories in the background of the near-extremal $AdS_6$. We obtained
the mass scale of the model by a comparison of the pion field action
with the usual Skyrme model action. We showed that the size of the
baryon is of order one in respect to the t' Hooft coupling but it is
still smaller than the scale of the dual QCD. So we introduced an
isospin 1/2 Dirac field for the five dimensional baryon and wrote an
effective action for it. Then, we reduced the five dimensional
action to the four dimensions using the mode expansion and obtained
the NN interaction in terms of the meson exchange potentials.

In our analysis such as the critical ones, the NN potential involves
only the pseudo-scalar, vector, and axial vector meson interactions.
In fact these meson exchange potentials play a special role in
producing the NN potential. The long-range part of the potential $(r
> 3fm)$ is mostly due to the one pion exchange mechanism which is
the strongest component among the isospin dependent components. In
the phenomenological potential models, the isospin dependent
pseudo-scalar meson exchange potential is of order $N_c^3$ while the
isospin independent component is of order $N_c$. So it is expected
that the $\eta$-meson exchange potential would be much weaker than
the $\pi$-meson exchange potential. We obtained $g_\pi=15.7$ and
$g_\eta=0.0$ at the leading order that is consistent with the
phenomenological interaction models.

Isospin independent scalar mesons are responsible for the attractive
interaction in the intermediate range ($(0.7 < r < 2fm)$) of the
potential. These components are the main reason for the nuclear
binding. Also in the phenomenological interaction models, the
strength of this interaction is equal to the vector meson exchange
with a minus sign. In fact, the radial shapes differ considerably at
short distances, ranging from the attractive to repulsive area. Some
of these phenomenological potential models involve only the scalar
meson exchange and the others contain the vector meson exchange
interaction term. We considered the vector meson exchange potential
in our analysis which produces the strong short-range repulsion. By
exchange the vector meson, $\rho$ can explains the small attractive
behavior of the odd-triplet state.

We compared our results with the available values of coupling
constants predicted in the four modern phenomenological interaction
models [Nijmegen(93), Paris, CD-Bonn, and AV 18 models] in Table
III. The remarkable point is that all of the meson-nucleon couplings
are calculated directly in our model whereas in the phenomenological
interaction models these values were obtained by fitting the NN
scattering data. It is also obvious from Table III that the values
of the coupling constants are widely different in the various
interaction models.

We believe that our non-critical holography model is more reliable
than the critical SS model to study the NN interactions for these
reasons:
\begin{itemize}
    \item Just like the SS model, there exist some KK modes which come from
the anti-periodic boundary conditions over the circle $S^1$. These
modes have the masses of the same order of magnitude as the lightest
glueballs of the four dimensional YM theory. The critical
holographic models such as the SS model, have some extra KK modes
too which do not belong to the spectrum of pure YM theory. These
undesired KK modes come from the extra internal space over which ten
dimensional string theory is compactified, for example, the $S^4$
sphere in the SS model. In the non-critical holographic model, which
we used here, there is no additional compactified sphere, so there
are no such extra KK modes and the QCD spectrum is clear from them.
Thus it seems that our model based on the non-critical holography is
much more reliable. As we mentioned earlier, we obtained the value
of $M_{KK}$ roughly half of its value in the SS model.

    \item  The size of the baryon is proportional to $1/\sqrt{\lambda
N_c}$ in the SS model and becomes zero at large 't Hooft coupling
while it is of order one in our calculations. So, our model does not
suffer from the zero size of the baryon.

    \item Also, the nucleon mass is obtained at $1.93\, M_{KK}$ in the SS model.
The mass scale which can describe the meson spectrum is roughly
$M_{KK}=940\,MeV$. So the nucleon mass is about $1.8$ GeV in the SS
model. It causes some inconsistency in analyzing the baryon, because
the scale of the system associated with the baryonic structure is
roughly half the one needed to fit to the mesonic data [36]. But in
our analysis the mass of the nucleon is obtained at roughly 920 MeV
which is very close to the neutron (proton) mass.

     \item In addition, our non-critical calculations can describe the
meson-nucleon couplings successfully at least at the leading order.
\end{itemize}

This holography potential model can be improved by considering the
contributions of derivative couplings and exchange of the other
mesons to the NN potential. Also, the couplings can be computed with
more accuracy however it seems that the contribution of heavy meson
exchange does not play a major role in such calculations. Moreover,
we can use the obtained potential here to study the nuclear
properties such as the NN scattering data and nuclear binding
energies. We leave them here as our future work.


\end{document}